\documentclass[%
 aip,
 amsmath,amssymb,
reprint,%
onecolumn,%
]{revtex4-1}

\usepackage{graphicx}% Include figure files
\usepackage{dcolumn}% Align table columns on decimal point
\usepackage{mathtools}
\usepackage{bm}% bold math
\usepackage[utf8]{inputenc}
\usepackage[T1]{fontenc}
\usepackage{mathptmx}
\usepackage{epstopdf}
\usepackage{lipsum} % For dummy text
\usepackage{fancyhdr} % Required for custom headers/footers
\usepackage{xcolor}

\newcommand{\ket}[1]{|#1\rangle}
\newcommand{\bra}[1]{\langle #1|}
\newcommand{\be}{\begin{eqnarray}}
\newcommand{\ee}{\end{eqnarray}}

\newcommand{\myfooter}{Preprint submitted to {\em Contributions to Plasma Physics}}

\fancypagestyle{plain}{
    \fancyhf{}
    \fancyfoot[C]{\thepage}
    \fancyfoot[L]{\footnotesize \myfooter}

}

\begin{document}

%\preprint{Submitted to Contributions to Plasma Physics}

\title[Improved Kelbg Potentials for $Z>1$ and Application to Carbon Plasmas]{Improved Kelbg Potentials for $Z>1$ and Application to Carbon Plasmas}
% Force line breaks with \\

\author{Heather D. Whitley}
 \email{whitley3@llnl.gov}
\affiliation{Lawrence Livermore National Laboratory, Livermore, California 94550, USA}%Lines break automatically or can be forced with \\
\author{Michael S. Murillo}
\affiliation{ 
Computational Mathematics, Science and Engineering, Michigan State University, East Lansing, Michigan 48824, USA
}%
\author{John I. Castor}%
\thanks{Deceased}
\affiliation{Lawrence Livermore National Laboratory, Livermore, California 94550, USA}
\author{Liam G. Stanton}
\affiliation{ 
Department of Mathematics and Statistics, San Jose State University, San Jose, California, USA
}%
\author{Lorin X. Benedict}
\affiliation{Lawrence Livermore National Laboratory, Livermore, California 94550, USA}
\author{Philip A. Sterne}
\affiliation{Lawrence Livermore National Laboratory, Livermore, California 94550, USA}%
\author{James N. Glosli}
\affiliation{Lawrence Livermore National Laboratory, Livermore, California 94550, USA}%
\author{Frank R. Graziani}
\affiliation{Lawrence Livermore National Laboratory, Livermore, California 94550, USA}%

\date{\today}% It is always \today, today,

\begin{abstract}
In this work, we present a general form for the electron-ion diffractive potential derived from the quantum pair density matrix and fit to the improved Kelbg potential for atomic numbers up to $Z = 54$. We apply classical molecular dynamics using the improved Kelbg potential for carbon with various forms of the Pauli potential to compute internal energies and pressures for hot, dense plasma conditions. Our results are compared to an equation of state model based on path integral Monte Carlo and density functional theory simulations to examine the extent to which the improved Kelbg potential reproduces the internal energy and pressure of carbon plasmas. The regions of validity for carbon agree generally with those derived previously for hydrogen once pressure ionization effects are incorporated. Based on our carbon results and previously published hydrogen studies, we discuss the general applicability and limitations of these potentials for equation of state studies in warm dense matter and high energy density plasmas. \end{abstract}

\maketitle

\section{Introduction}
\label{Intro}

In the early 2000s, motivated by uncertainties in the equation of state and plasma transport properties of warm dense matter and hot, dense plasmas relevant to inertial confinement fusion (ICF) 
experiments at the 
National Ignition Facility (NIF), the Cimarron Collaboration\cite{hedp12} was formed under the Laboratory Directed Research and Development program at LLNL to develop a molecular dynamics (MD) capability that allowed us to the investigate the micro-physics of non-equilibrium, hot, dense plasmas, relevant to ICF.  Key properties of interest for Cimarron included electron-ion coupling,\cite{benedict2009molecular,Benedict12,benedict2017molecular} 
electron and ion transport processes,\cite{ellis2011studies,Grabowski2013,hau2013nonequilibrium,haxhimali2014diffusivity,Whitley15,hau2015electric,hau2015reproducible,haxhimali2015shear,Desjarlais17,mackay2017species,grabowski2020review} and 
the impacts of many-body physics on thermonuclear burn.\cite{hau2013modeling,whitley2015localization}
A fundamental 
question of this effort was understanding the fidelity to which effective interaction potentials could be applied within the ddcMD code\cite{fattebert2012dynamic} to study plasmas.  As the temperature of a plasma increases, the 
statistical thermodynamics of the many body system approaches the ideal gas limit, however, quantum mechanics remains important in the close inter-particle collisional dynamics.  At small
electron-ion distances the quantum nature of the electrons prevents the so-called "Coulomb catastrophe,'' where electrons effectively spiral into the nucleus, while for electron-electron 
collisions Fermi statistics remain important to the overall determination of the collisional dynamics.\cite{DeWitt69}  
One way of incorporating the quantum nature of the electrons into a classical simulation is to modify the normal Coulomb potential in order to mimic quantum behavior using a quantum statistical potential.  The derivation of 
quantum statistical potentials (QSPs) or "effective potentials'' to describe plasmas has a long history, going back almost a century.\cite{UG,Lado,DunnBroy,Kelbg,Deutsch78a,minoo,filinov03,murillojones,bonitz2023gunter}  The primary premise 
of these studies is similar to the underlying theory of the quantum path integral Monte Carlo (PIMC) method, namely that a quantum many-body system can be effectively mapped to a system of 
classical particles within a set of controlled approximations.  
%REV22626
The term "QSP" arises due to the fact that the potentials are rooted in the description of a quantum statistical mechanical ensemble.  In general, these potentials 
are derived from an exact solution to the pair problem for the quantum thermal ensemble.  In the case of interacting fermions, approximations are made 
to account for antisymmetry.

We have previously reported the application of QSPs to studies of electron-ion relaxation and conductivity in hydrogen plasmas\cite{hedp12,Benedict12,Whitley15}, and several authors have recently published 
reviews of the QSP approach, as well as detailed  
studies of hydrogen plasmas within the QSP approximation as implemented in both molecular dynamics and path integral Monte Carlo simulations.\cite{Filinov04,bonitz2023gunter,Filinov23}
In this paper, we firstly describe the path integral representation of the quantum partition function and use this as a basis upon which to generally describe the approximations that are applied within the 
quantum statistical interaction potentials and the molecular dynamics approach. We apply the well-known pair density approximation to 
 compute the exact pair density matrix for a variety of electron-ion pairs, with atomic number $Z=1$ (hydrogen) to 54 (xenon).  We numerically fit an updated form of the Pad\'{e} approximant, originally 
described by Filinov {\em et al.} for electron-proton interactions, that provides a reasonable fit of the improved Kelbg potential for all values of $Z$ considered in this study, 
thusly providing a general form for the improved Kelbg potential for fully ionized species beyond hydrogen.  
%REV22626
While many modern MD codes allow for the inclusion of tabulated forms of interaction potentials, the analytic form here provides a generalized description for any value of the nuclear charge state, eliminating the need 
to explicitly recalculate the pair density matrix for each species of interest.  
Analytic forms also allow for greater computational efficiency and smaller memory footprint by eliminating the need for tabular lookups over the course of a simulation.

Due to its applicability as an ablator material in many ICF experiments, 
%REV22626
prevalence in astrophysical objects, and many applications in both scientific and industrial processes, carbon and carbon-containing compounds have been the topic of many theoretical and experimental studies over the last two
decades.\cite{batani2004hugoniot,Nagao2006,knudson2008shock,PhysRevLett.108.115502,Hamel12,Benedict14,Sanchez15,kraus2015complex,Hu2016,Zhang2017,Zhang2018CH,gaffneyreview,Doppner18,PhysRevE.98.043204,Millot20,Kritcher20,Zhang2020,vorberger2020structure,Lazicki21,knudson2021interplay,Swift22,Hu2022,ao2023exploring,Oleynik2024}
This broad interest, as well as the availability of an EOS model based on quantum simulations with which to compare, inspired us to apply the improved Kelbg potential to ddcMD simulations of carbon 
plasmas as a test of the application of this method for $Z>1$. 
We examine the validity of the QSPs for carbon by comparing the simulated internal energies and pressure for a range of conditions to a previously published equation of state table\cite{Benedict14}  
%REV22626
that is based on quantum density functional theory and restricted PIMC simulations.  We discuss the validity of the QSP approach within the context of the average occupation of the K-shell, as calculated by the 
Purgatorio code\cite{Wilson06}.  We further relate the results of this study to the earlier work of Barker,\cite{Barker71} which focused on the derivation of QSPs for hydrogen, and discuss 
the validity of the application of QSPs to the calculation of thermodynamic properties.

\section{Quantum Statistical Potentials}
\label{QSP}
%\section{Finite temperature path integral Monte Carlo}

Jones and Murillo\cite{murillojones} reviewed in detail the derivation of various forms of the quantum statistical potentials, and recently Bonitz {\em et al.} published a 
full review of the Kelbg and improved Kelbg interaction potential that we focus on in this study.\cite{bonitz2023gunter}  Here, we give a brief overview of the 
quantum statistical potential (QSP) approach and how potentials are derived from the quantum pair density matrix.  We also discuss 
assumptions involved in the 
 application of the QSPs.  In general, the QSP approach trades the accuracy of the fully quantum treatment for a pair potential that is modified by two-body quantum physics that can be employed 
in classical molecular dynamics (MD) simulations.  The QSPs can also be used in classical Monte Carlo (MC) and 
%REV22626
integral equation theoretical approaches, such as hypernetted chain, thusly offering a 
substantially more efficient means of computing the thermodynamic properties of a plasma in regions where the necessary approximations are valid (in general, for fully ionized systems in which 
the electron density is low enough to avoid the need for explicit inclusion of three-body or higher quantum correlations.)

Starting from the definition of the partition function, $Z\equiv \mathrm{Tr}(\hat{\rho})$, where $\hat{\rho}=e^{-\beta\hat{H}}$ and $\beta = 1/{k_BT}$,\cite{feynman53,feynman72}    
the density matrix can be written as a function of the coordinates of all particles in the system, $\mathbf{R}$, 
\begin{eqnarray}
\rho(\mathbf{R},\mathbf{R}^\prime;\beta)&=&\bra{\mathbf{R}}e^{-\beta\hat{H}}\ket{\mathbf{R}^\prime}\nonumber\\
&=&\sum_s \Psi_s^\ast(\mathbf{R})\Psi_s(\mathbf{R}^\prime)e^{-\beta E_s},
\label{rho1}
\end{eqnarray}
where $\{\ket{\Psi_s}\}$ and $\{E_s\}$ are the eigenstates and respective eigenvalues of the many body Hamiltonian,
\begin{eqnarray}
\hat{H}&\equiv& \hat{T} + \hat{V} = \sum_{i=1}^N \left[ -\frac{\hbar^2}{2m}\vec{\nabla_i}^2 + \sum_{j<i} U_{ij}(\hat{r}_{ij})\right].
\label{ham}
\end{eqnarray}
Here, we have assumed that the potential energy can be decomposed into a pairwise sum over all $N$ particles in the system.  
It is apparent that the density matrix represents the probability of a system moving from configuration 
$\mathbf{R}$ to configuration $\mathbf{R}^\prime$, as defined by the density of states and temperature of the system.  
With this definition of the density matrix, the partition function is a configuration space integral, $Z=\int d\mathbf{R}\rho(\mathbf{R},\mathbf{R};\beta)$ and 
the expectation value of any operator ($\hat{O}$) in a system at finite temperature is given by
\begin{equation}
\langle \hat{O} \rangle = \frac{1}{Z} \int d\mathbf{R} d\mathbf{R}^{\prime} \bra{\mathbf{R}^{\prime}} \hat{O} \ket{\mathbf{R}} \rho(\mathbf{R},\mathbf{R}^{\prime};\beta).
\label{expect1}
\end{equation}
 This expression of the expectation value as a configuration space average over a probability distribution is analogous to the definition of the average value of a classical observable 
 over a classical probability distribution, which allows the quantum system to be mapped onto a classical system.  A corresponding equation for the classical $N$-body system is obtained by
multiplying the partition function by a particular form of unity~\cite{murillojones},
\begin{equation}\begin{split}
Q&=\frac{C}{C}\int
d\mathbf{R}\rho(\mathbf{R},\mathbf{R};\beta)\\
&=\int
\frac{d^{3N}p}{(2\pi\hbar)^{3N}N!}e^{\sum_i \frac{-\beta p_i^2}{2m}}\int
d\mathbf{R} \frac{\rho(\mathbf{R},\mathbf{R};\beta)}{C}\;.
\label{semiclass}
\end{split}
\end{equation}
The ability to perform this mapping arises from the fact that the 
quantum partition function is identical to a classical partition function where the total number of particles is multiplied by the number 
of discretized configurational states of the quantum system.  Therefore, there is an exact procedure for understanding many of the 
properties of a quantum system in terms of classical statistical mechanics.\cite{chandler81}  
 
At high temperatures, the density matrix is often approximated by making use of 
the pair product approximation, where the exact pair density matrix is used to construct the potential energy terms of the high temperature density matrix:\cite{Barker79,pollock84,ceperley95}  
\begin{equation}
\rho(\mathbf{R},\mathbf{R}^\prime;\beta)\approx 
\left[\prod_{i=1}^N\rho_F(\mathbf{r}_i,\mathbf{r}_i^\prime;\beta)\right]
\left[\prod_{i< j}\frac{ \rho_2(\mathbf{r}_{ij},\mathbf{r}_{ij}^\prime;\beta)}%
{\rho_F(\mathbf{r}_{ij},\mathbf{r}_{ij}^\prime;\beta)}\right],
\end{equation}
where
\begin{equation}
\rho_F(\mathbf{r}_i,\mathbf{r}_i^\prime;\beta)=
\left(\frac{2\pi \beta\hbar^2}{m_i}\right)^{-3/2} 
\exp\left[-\frac{m_i|\mathbf{r}_i-\mathbf{r}_i^\prime|^2}{2\hbar^2\beta}\right]
\end{equation}
is the free particle density matrix and
$\rho_2(\mathbf{r}_{ij},\mathbf{r}_{ij}^\prime;\beta)/%
\rho_F(\mathbf{r}_{ij},\mathbf{r}_{ij}^\prime;\beta)$
is the interaction portion of the pair density matrix for each pair of
particles $ij$.
This method is particularly useful when the potential energy can be written as a pairwise sum, as in Eq.~\ref{ham}.  The pair approximation here is analogous to that of PIMC, except that in the 
classical limit, the off-diagonal portions of the pair density matrix 
are assumed to be zero (ie. quantum non-local exchange is neglected).  In so doing, the QSPs derived from the 
pair approximation are only valid for plasmas in which three-body or higher order interactions can be accurately represented as a pairwise sum, which is only strictly true for situations where the 
electrons remain relatively weakly coupled and non-degenerate, and the plasma is nearly fully ionized.  

In the case of a fully ionized plasma, the calculation of 
a quantum statistical potential within the pair approximation is straightforward.\cite{Barker71} 
%note: add analogy of pair approximation to pair approximation of PIMC in the appropriate section
By assuming the pair product decomposition of the density matrix, a pairwise QSP is defined as
\begin{equation}
U(r_{ij},\beta)=-\frac{1}{\beta}
\log\left[{\frac{\rho_2(\mathbf{r}_{ij},\mathbf{r}_{ij};\beta)}%
{\rho_F(\mathbf{r}_{ij},\mathbf{r}_{ij};\beta)}}\right],
\label{rhopot}
\end{equation}
where $\rho_2(\mathbf{r}_{ij},\mathbf{r}_{ij};\beta)$ is a diagonal element of the exact pair density matrix written in the position basis.  
The diagonal part of
$\rho_F(\mathbf{r}_{ij},\mathbf{r}_{ij};\beta)$ is a constant,
$\left(2\pi \beta\hbar^2/\mu_{ij}\right)^{-3/2}$, where $\mu_{ij}$ is
the reduced mass for the pair of particles.  For a two-body (pair)
problem, the Schr\"{o}dinger equation can be solved
exactly and $\rho_2$ is evaluated according to the Slater sum,
\begin{equation}
\rho_2(\mathbf{r_{ij}},\mathbf{r_{ij}}^\prime;\beta)=\sum_s\Psi_s(\mathbf{r_{ij}})
e^{-\beta E_s}\Psi^\ast_s(\mathbf{r_{ij}}^\prime).
\label{denm}
\end{equation}
Since all $\{\Psi_s\}$ and $\{E_s\}$ can be calculated analytically for 
the pair problem, evaluation of $\rho_2$ is straightforward, and 
quantum exchange symmetry can be enforced through the basis functions $\{\Psi_s\}$.  
The main numerical difficulty in computing Eq.~\ref{denm} 
comes from the fact that at very high temperatures the number of
states that must be included in the sum becomes prohibitively large.
For Coulomb systems, an efficient method for calculating the pair
density matrix was developed by Pollock.~\cite{Pollockpaper} 
The approximation that has been 
introduced at this point is the neglect of quantum mechanical nonlocal exchange, which mathematically corresponds to the off-diagonal terms of the density matrix.  

Barker\cite{Barker71} and others \cite{DunnBroy,murillojones,bonitz2023gunter} have 
outlined detailed regions 
of validity for the application of QSPs to hydrogen plasmas based on theoretical considerations.  The general regions of validity correspond to 50\% or higher ionization of the K-shell and density-temperature conditions that avoid the need for the inclusion of three-body or higher interactions.  Eq. 10 of Ref.~\onlinecite{Barker71} (shown below) and Eq. 65 of 
Ref.~\onlinecite{murillojones} provided very similar functional fits to the condition under which three-body terms can be neglected that was originally derived by Dunn and Broyles,\cite{DunnBroy}
\be
\label{validline}
\mathrm{log}_{10}(n_e)\lesssim \frac{9}{4}\mathrm{log}_{10}(T)+11.6,
\ee
where $n_e$ is the total electron density in cm$^{-3}$ and $T$ is the temperature in Kelvin.  In principle, applying a many-body simulation method, like classical MD or classical Monte Carlo, would 
incorporate missing three-body effects at the classical level, hence Jones and Murillo noted that explicit comparisons of thermodynamic quantities derived from full many-body simulations using the 
QSPs would be required to fully investigate their validity for computing various physical properties.\cite{murillojones}

Historically, most derivations of an effective QSP for the Coulomb pair problem made use of perturbation theory\cite{Kelbg,DunnBroy} or 
neglected contributions from the atomic orbitals completely.\cite{Deutsch78a,Deutsch78b}  To our knowledge, Barker was the first to derive a quantum interaction potential for hydrogen from the full Slater sum.\cite{Barker71}  
The more approximate treatments that do not include the full Slater sum 
have since been shown to differ more substantially from fully quantum treatments of hydrogen plasmas.\cite{Filinov04,hedp12,Benedict12,Whitley15,Filinov23,bonitz2023gunter}  
In the work of Filinov et.al,\cite{filinov03,Filinov04,bonitz2023gunter} an analytic equation is introduced as an improvement of the 
potential originally derived by Kelbg.  In this work, a fitting parameter 
($\gamma_{ij}$) based on the limiting value of Eq.~\ref{denm} including the atomic orbitals at $r_{ij}=0$ 
is introduced to the Kelbg potential:
\begin{widetext}
\begin{equation}
U(r_{ij},\beta)=\frac{q_i q_j}{r_{ij}}\Biggl[1-e^{-\left(\frac{r_{ij}}{\lambda_{ij}}\right)^2}+\sqrt{\pi}\frac{r_{ij}}{\lambda_{ij}\gamma_{ij}}\left(1-\mathrm{erf}\left[\gamma_{ij}\frac{r_{ij}}{\lambda_{ij}} \right]\right)\Biggr]\;.
\label{kelbgpot}
\end{equation}
\end{widetext}
Here, $i$ and $j$ denote the indices of each particle, $q_i$ and $q_j$ are the charges of the particles, $\lambda_{ij}$ is the thermal de Br\"{o}glie wavelength ($\lambda_{ij}^2=\frac{\hbar^2\beta}{2\mu_{ij}}$), and $\gamma_{ij}$ is a temperature-dependent fitting parameter. We note that
%REV22626
in principle, one can also choose to treat the distance associated with quantum effects as a temperature dependent fitting parameter, $\lambda^{\prime}$, which replaces the thermal de Br\"{o}glie wavelength in Eq.~\ref{kelbgpot}.\cite{Benedict12}  
%REV22626
In the weak coupling limit at high temperatures, $\gamma_{ij}= 1$ and 
 the usual value of the thermal de Br\"{o}glie wavelength provides good agreement between Eq.~\ref{kelbgpot} and the potential derived from the exact pair density matrix.  
At low temperatures, 
the value of $\gamma_{ij}$ decreases for the electron-ion interaction and increases for the electron-electron interaction, 
consistent with Ref.~\onlinecite{Filinov04}.  Invoking $\lambda^{\prime}$ as a fitting parameter can improve the agreement between the analytic form for the improved Kelbg potential and the numerically 
determined quantum pair potential for $r>0$ in the case of the attractive interactions.  
We find that the deviation of $\lambda^{\prime}$ from the thermal de Br\"{o}glie wavelength for the interacting pair increases as the bound state contribution to the Slater sum increases, 
and in fact, this deviation is qualitatively indicative of the transition between a high temperature regime, where the thermal de B\"{o}glie wavelength is small compared to quantum mechanical 
average radius of the electron around the nucleus, 
to a lower temperature regime, where the average electron radius is similar to the thermal de Br\"{o}glie wavelength and a more accurate quantum description is needed.  In 
this work, we have thusly abandoned the use of $\lambda^{\prime}$, noting that if this sort of modification of the improved Kelbg 
form is needed to fit the quantum statistical potential, then the temperature is too low for the QSP treatment to be adequate 
or accurate in comparison to a fully quantum method.  A parametrized fit for $\gamma_{ei}$ (explicitly corresponding to an electron-ion interaction) will be presented in the next section.   In practice, the large mass of the 
ionic species leads to very small values of $\lambda_{ij}$ and $\gamma_{ij}$ is essentially equal to 1 for any pair of ions.  The Kelbg potential converges to the classical Coulomb potential in this limit.

In this study and in our previous work on hydrogen plasmas,\cite{hedp12, Benedict12,Whitley15} we computed the exact pair density matrix as a function of $r{_{ij}}$ for each pair of species in the problem neglecting exchange symmetry, then fit the resulting 
potentials to Eq.~\ref{kelbgpot}.  We approximately incorporated the impact of fermion exchange on the electrons by applying a Pauli potential.  Explicitly, rather than including the exchange symmetry in the calculation of the pair density matrix used to define the fitted improved Kelbg potential, the electron-electron interaction in our studies includes a separate Pauli term:
\begin{eqnarray}
U_{ee}(r_{ij},\beta) &=& U(r_{ij},\beta)  + U^{P}(r_{ij},\beta)
%V(\mathbf{R}) &=& \sum_{i=1}^{N_i} \left[\sum_{j<i} U_{ij}(r_{ij}) + \sum_{k=1}^{N_e} U_{diff}(r_{ik})\right] \\
%& & + \sum_{k=1}^{N_e} \sum_{l < k} \left[ U_{diff}(r_{kl}) + U_{Pauli}(r_{kl})\right] 
\end{eqnarray}
Here, $U_{ee}(r_{ij},\beta)$ is assumed to take the form of Eq.~\ref{kelbgpot} with $\gamma_{ee}$ (explicitly defined for an electron-electron interaction) is  derived by fitting to calculations of the exact pair potential. 
Using the Pollock algorithm to compute the exact pair potentials, we find
\begin{equation}
\label{gamee}
\gamma_{ee}=1+\frac{0.0321}{(k_BT)^{0.4664}},
\end{equation}
where $k_BT$ is in Hartrees, as previously reported.\cite{Benedict12}
We have used two forms of the Pauli potential ($U^P(r_{ij},\beta)$) that were derived by previous authors for non-interacting fermions.\cite{UG,Deutsch78a,Deutsch78b,Lado,DunnBroy}  Lado\cite{Lado}  derived a density dependent form of the effective Pauli potential for non-interacting 
fermions starting from the path integral expression for the full $N$-body partition function.  By applying the Born-Green hierarchy, Lado arrived at a spin-dependent, pair-wise form for the interaction in the limit of low density:
\begin{eqnarray}
U^P_{Lado,s}(r_{ij},\beta)\approx -\beta^{-1}\ln\left[1 \pm \frac{\mathrm{exp}(-2\pi r^2/\Lambda^2)}{2s+1}  \right],
\label{ladopot}
\end{eqnarray}
where $\Lambda = \sqrt{ \frac{2\pi \hbar^2}{\beta m} }$.
This form of the Pauli interaction is similar to applying the "free-particle" fixed-node approximation in quantum Monte Carlo methods; it is exact in the limit of non-interacting fermions and breaks down as two-body and higher interactions become 
important.  The efficacy of the fixed-node approximation as a function of temperature in PIMC calculations has been studied in detail by several authors through comparisons with density functional theory and experimental data,\cite{Driver2012,Zhang2017,Zhang2018CH,Doppner18,Kritcher20,Zhang2018,Zhang2019,Zhang2020,militzerAl}
 and some authors have extended the applicability of PIMC to lower temperatures.\cite{Driver2012,DH18,BD20,TD25} In general, the limits of applicability of the Lado potential in classical simulations as a function of temperature are expected to be similar to those 
 found in PIMC, with PIMC having greater accuracy in quantum, strongly interacting regimes due to the inclusion of the full density matrix description rather than just the diagonal components in the pair-wise interaction scheme. Jones and Murillo\cite{murillojones} subsequently derived a parametric fit for the spin-averaged version of the Lado potential: 
\begin{equation}
\label{Lado2}
U^P_{Lado}(r_{ij},\beta) = -\beta^{-1} \ln \left[ 1 - \frac{1}{2} A(\tau) \exp\left( -2\pi B(\tau) \frac{r^2}{\Lambda^2} \right) \right],
\end{equation}
%REV22626
where $\tau={(\beta T_F)}^{-1}=T/{T_F}$ is the degeneracy parameter,
\begin{equation}
A(\tau) = 1 + \frac{a_1}{1 + a_2 \tau^{a_3}}, \quad B(\tau) = 1 + b_1 \exp\left( -b_2 \tau^{b_3} \right) / \tau^{b_4},
\end{equation}
and the parameters are $a_1 = 0.2975$, $a_2 = 6.090$, $a_3 = 1.541$,$b_1 = 0.0842$, $b_2 = 0.1027$, $b_3 = 1.096$, and $b_4 = 1.359$.  This version of the Lado potential is implemented in ddcMD and converges with the Uhlenbeck and Gropper\cite{UG} potential at high temperatures.  
 In addition to the Lado potential, our ddcMD studies\cite{hedp12} have commonly applied a formulation of the Pauli potential originally derived by Deutsch, Minoo, and Gombert\cite{Deutsch78a,Deutsch78b} in the form published by Hansen and MacDonald\cite{HansMcD81}:
 \begin{equation}
\label{HansenPauli}
U^{P}_{Deutsch}(r) = \beta^{-1}\ln(2)\exp\left( -[\pi\ln(2)]^{-1} r^2/\Lambda_H^2\right),
\end{equation}
where $\Lambda_{H} = \hbar/\sqrt{\pi m_eT}$.
 We note that it is also possible to account for fermionic exchange symmetry explicitly in the calculation of the electron pair density matrix.  This approach was applied in the work of Filinov 
 {\em et al.}\cite{Filinov04} and Bonitz {\em et al.}~\cite{bonitz2023gunter}, wherein an empirically fit analytic form for the full e-e interaction potential with exchange symmetry included was published.
 
%REV22626 
  The Deutsch and the Lado potentials are essentially the same in the high temperature limit,\cite{murillojones} we have therefore used the Deutsch potential in the higher temperature calculations 
 in this study, primarily for conditions corresponding to $T/T_F > 10$.  We make some comparisons between simulations with a Pauli potential to simulations at the same temperatures and densities without a Pauli potential.  It is also worth noting that classical linear response theory can also be used to derive the effective interaction potential between two particles,\cite{GV05,Kuk79,Kuk21} however potentials derived in this 
 manner already account for the screening due to many-body interactions and are not appropriate for use in explicit MD or MC simulations like the ones described here.

\subsection{The Improved Kelbg Potential for $Z>1$}

We employed the Slater sum approach described by Pollock~\cite{Pollockpaper}, as well as its implementation within the universal path integral (UPI) code coupled with matrix squaring~\cite{ceperley95,storer68,storer73}, to compute pair density matrices for electron-ion systems with nuclear charge 
%REV22626
numbers ($Z$) ranging from 1 to 54.  This range of $Z$ is largely representative of states accessible in laboratory high energy density physics experiments, 
where temperatures are  $\sim$1-10~keV in the DT fusion fuel and $\sim$~300~eV in the gold hohlraum,\cite{Aybar25} however the analytic fit for the potentials derived here likely holds for any value of $Z$.
 Quantum statistical potentials (QSPs) were subsequently extracted using Eq.~\ref{rhopot}. The resulting potentials for hydrogen are presented in Figure~2 of Ref.~\onlinecite{hedp12} and are not reproduced here.

\begin{figure*}
\centering
a) \rotatebox{-90}{\resizebox{2.5in}{!}{\includegraphics{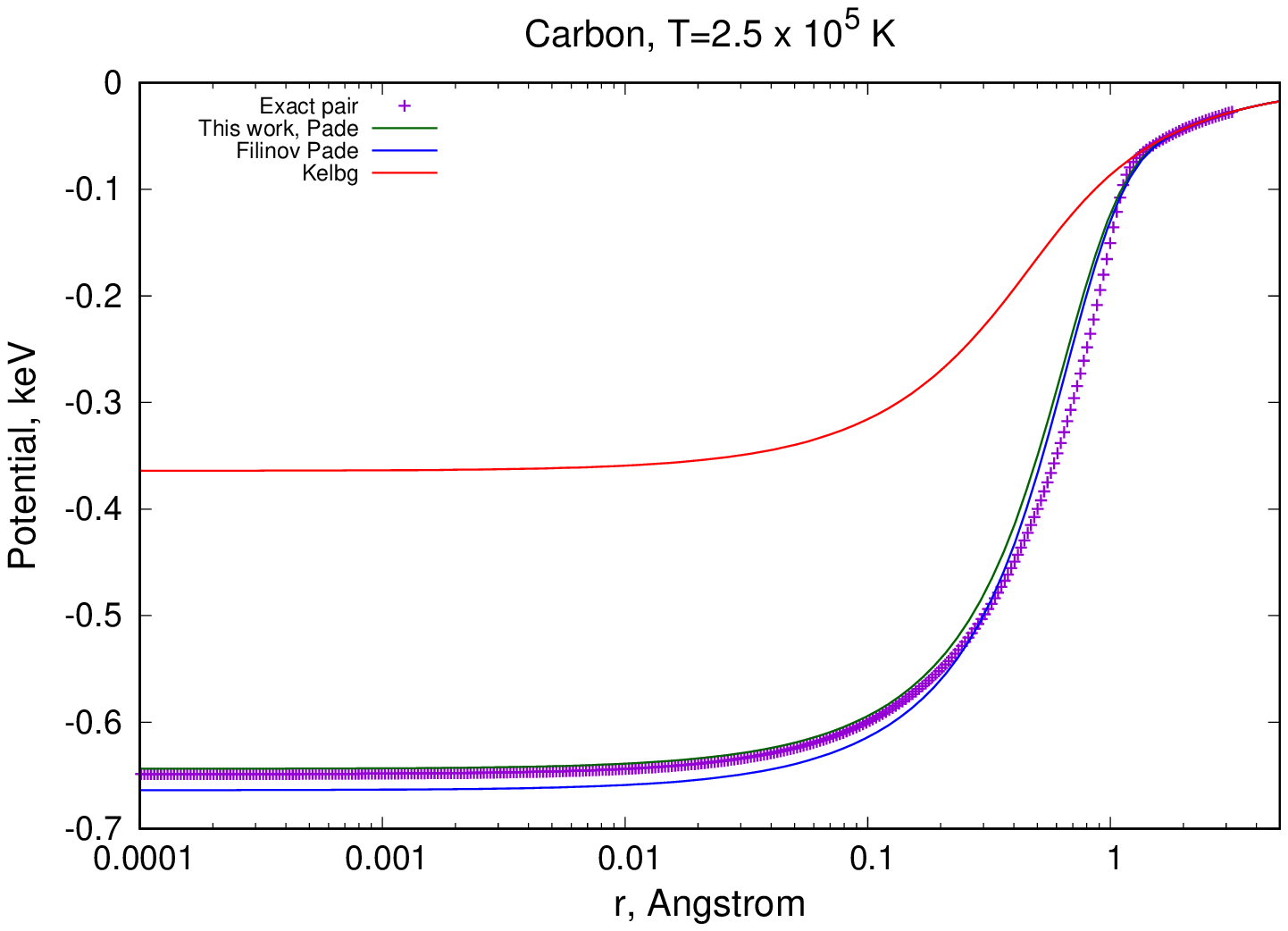}}}

b) \rotatebox{-90}{\resizebox{2.5in}{!}{\includegraphics{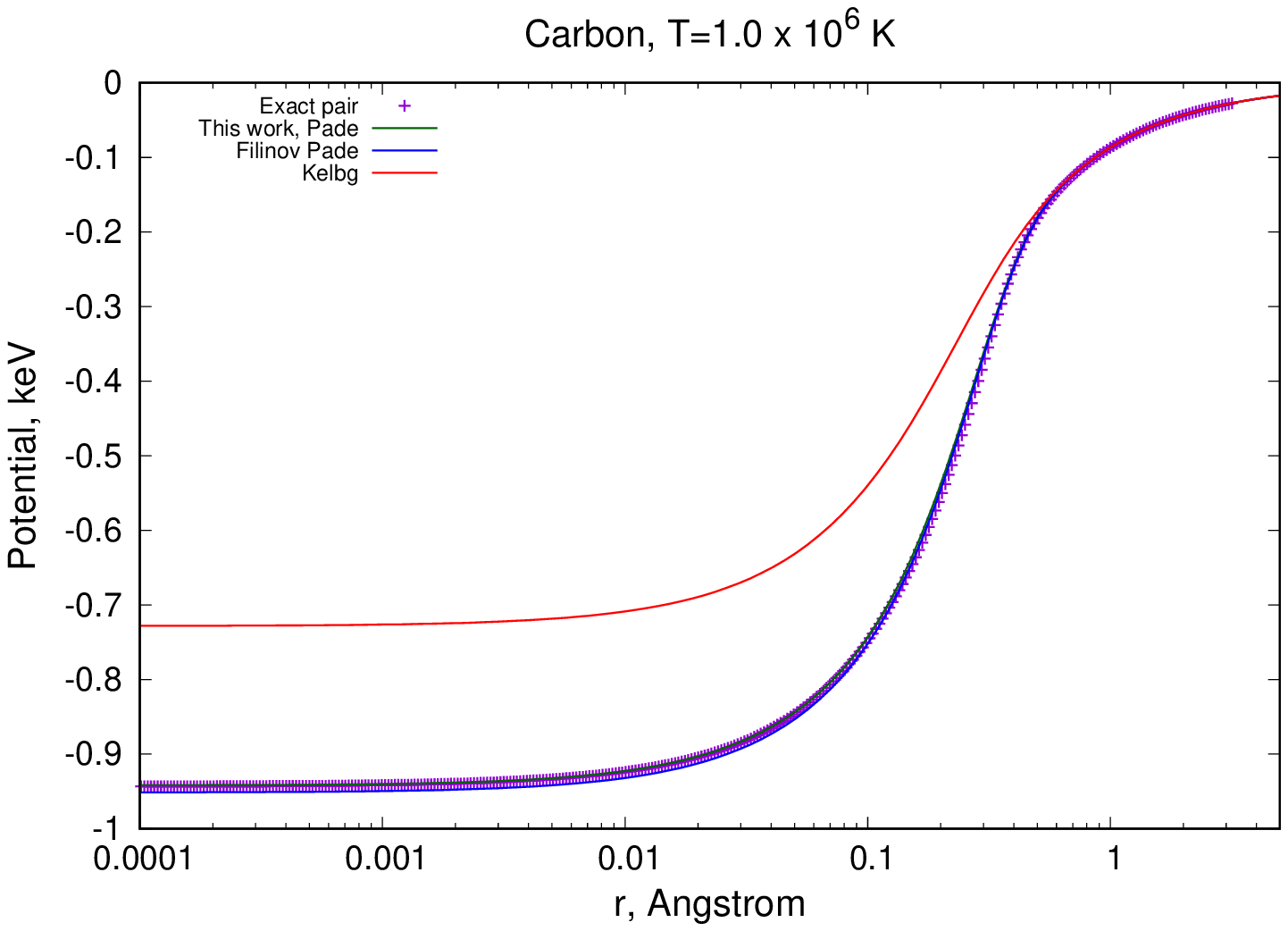}}}
\caption{Diffractive potentials for the electron--carbon ion interaction at $2.5\times 10^5$~K (a) and $1.0\times 10^6$~K (b). Potentials computed from the exact pair density matrix are shown as purple symbols. The red line denotes the Kelbg potential~\cite{Kelbg}, while the green and blue lines correspond to the improved Kelbg potential using the Pad\'{e} approximant of this work and of Filinov {\em et al.}~\cite{Filinov04}, respectively.}
\label{carbonpots2}
\end{figure*}

Figure~\ref{carbonpots2} shows representative QSPs for carbon at temperatures of $2.5\times 10^5$~K and $1.0\times 10^6$~K. Results from the exact pair density matrix calculations are indicated by purple symbols. The green lines correspond to our fitted expressions for the potentials based on Eq.~\ref{kelbgpot}, as described below, and the red lines correspond to the traditional ("unimproved") Kelbg potential. Both 
figures demonstrate the importance of including the bound state contributions to the potential as the traditional Kelbg potential indicates significantly weaker attractive interactions at small radius than the exact pair potential.  
At lower temperatures, where the K-shell of the isolated electron-ion pair is predominantly occupied, the form of the QSP is also significantly modified at distances exceeding the thermal de~Broglie wavelength.  
This is visible as the small discontinuity in the exact pair potential at $T=2.5\times 10^5$~K. This discontinuity arises in the pair potential due to the strong contribution of bound states and the ionization gap between the highest occupied bound states and the continuum for the pair problem.  Fig.~\ref{carbonpots2} also shows a version of the improved Kelbg potential based on Ref.~\onlinecite{Filinov04}, as described in more detail below.  

In the present study, we determined the $\gamma_{ie}$ parameter of the improved Kelbg potential for $Z$ values from 1 to 54 by applying a least-squares fitting procedure (implemented in gnuplot) to each exact pair potential using Eq.~\ref{kelbgpot}, with $\gamma_{ei}$ as a free parameter. The fitted $\gamma_{ei}$ values were then parameterized using a generalized Pad\'{e} approximant to second order in the dimensionless parameter $x\equiv x(T,Z)$, defined as
\begin{eqnarray}
x(T,Z)=\sqrt{\frac{8\pi k_BT}{Z^2\mathrm{Ha}}},
\label{xpade}
\end{eqnarray}
where $\mathrm{Ha}$ is the Hartree energy in units of $k_BT$.  The resulting Pad\'{e} fit as a function of $Z$ and $T$ is given by
\begin{eqnarray}
\gamma_{ei}(T,Z)=\frac{0.85x+2x^2}{1+1.3x+2x^2}.
\label{newpade}
\end{eqnarray}
The fit to a Pad\'{e} approximant was inspired by the work of Ref.~\onlinecite{Filinov04}, wherein the authors derived a fitted Pad\'{e} approximant for the interaction between an electron and a proton.  
The quantity $x(T,Z)$ effectively corresponds to a measure of the thermal energy to the ground state energy of a single electron atom of charge $Z$.  It provides a smooth functional form for capturing the transition from the weak-coupling limit ($\gamma_{ei}=1$), where perturbation theory holds and the original Kelbg potential is regained, to the strong-coupling limit ($\gamma_{ei}=0$), where 
bound state contributions to the thermal density matrix dominate and the QSP approach breaks down.

In order to compare our form for the QSPs with Ref.~\onlinecite{Filinov04}, we used our definition for $x(T,Z)$ (Eq.~\ref{xpade}) in the Pad\'{e} approximate of Ref.~\onlinecite{Filinov04} to compute the carbon potentials
using Eq.~\ref{kelbgpot} which explicitly accounts for the value of $Z$.
  The results are shown as blue lines in Fig.~\ref{carbonpots2}.  Comparison of the exact pair potential obtained from the Slater sum with the Pad\'{e} approximant for carbon at $T=2.5\times10^5$~K exemplifies a trend that is observed for all $Z$: the values that we derived for the exact pair potential at $r=0$ are less attractive than those predicted by the Pad\'{e} fit from Ref.~\onlinecite{Filinov04}. This discrepancy is most pronounced at low temperatures, but persists at higher temperatures, as shown for carbon at $T=1\times 10^6$~K. These observations motivated the development of a modified Pad\'{e} approximant to improve agreement with our results from the pair density matrix, 
though ultimately the differences are small and potentially related to small numerical differences in our calculations of the exact pair potential as compared to Ref.~\onlinecite{Filinov04}.

\begin{figure*}
\centering
a) \rotatebox{-90}{\resizebox{2.5in}{!}{\includegraphics{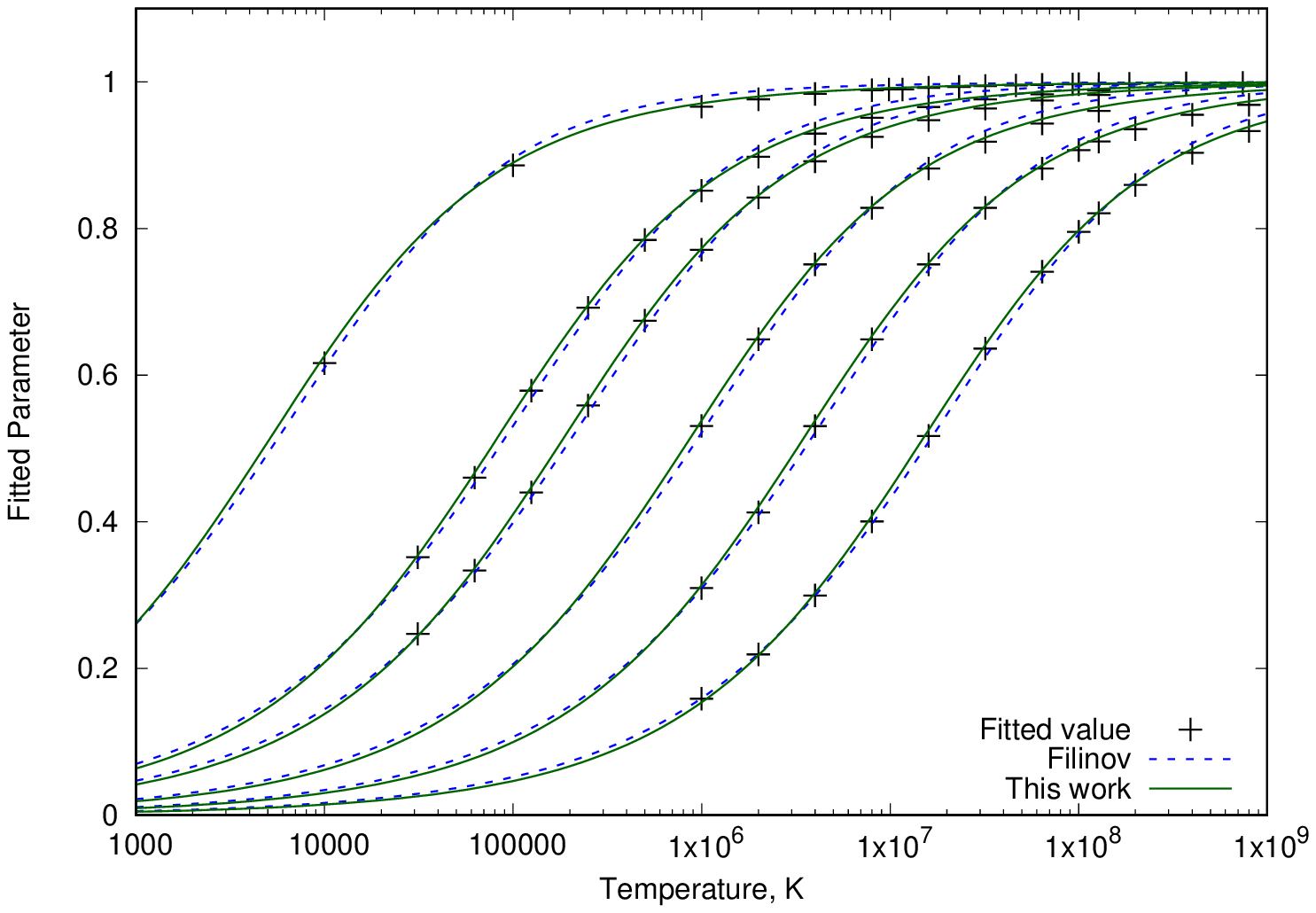}}}

b) \rotatebox{-90}{\resizebox{2.5in}{!}{\includegraphics{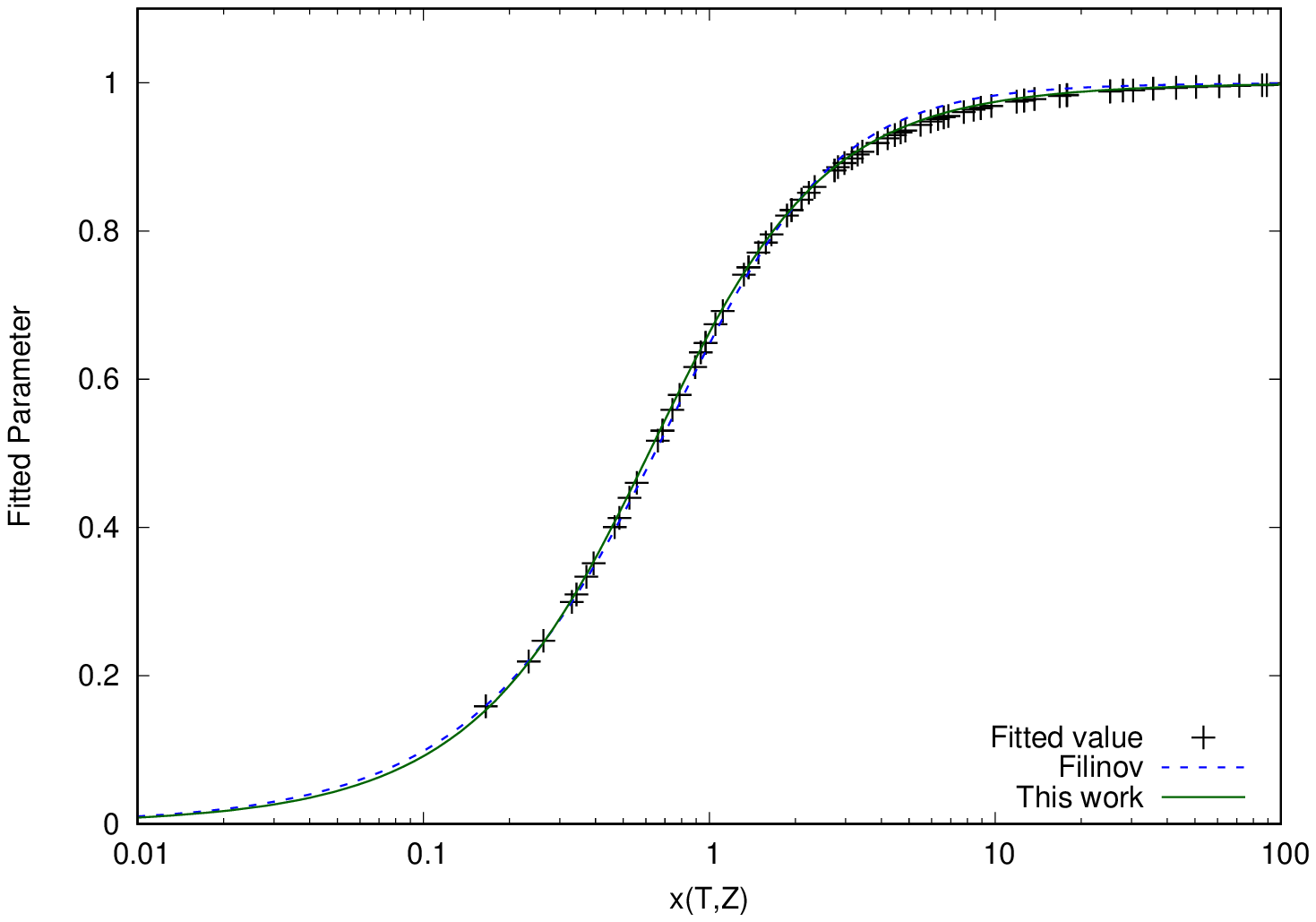}}}
\caption{Fitted parameter $\gamma_{ei}$ as a function of temperature (a) and as a function of the parameter $x(T,Z)=\sqrt{8\pi k_B T/(Z^2\mathrm{Ha})}$ (b).  Figure a) shows the fitted values from the exact Slater sum (crosses), the Pad\'{e} approximant of Filinov {\em et al.}~\cite{Filinov04} (blue dashed lines), and the present work (green lines). Curves correspond to atomic numbers $Z=1$, 4, 6, 13, 26, and 54, from left to right. Figure b) shows the fitted parameter as a function of $x(T,Z)$ for the same $Z$ values, illustrating the utility of $x(T,Z)$ for fitting $\gamma_{ei}$.}
\label{approximants}
\end{figure*}

Figure~\ref{approximants} presents the fitted parameter $\gamma_{ei}$ for a range of ionic species. Black crosses denote values obtained from numerical fits to the QSP computed from the Slater sum for the electron-ion Hamiltonian, while the blue and green lines correspond to the Pad\'{e} approximants of Ref.~\onlinecite{Filinov04} and Eqs.~\ref{xpade} and \ref{newpade}, respectively.  The most significant difference between our updated parameterization for $\gamma_{ei}$ and the expression given in Ref.~\onlinecite{Filinov04} is that our values do not approach $\gamma_{ei}=1$ as rapidly for increasing temperature.  The fit presented here is valid for all $Z$ considered in this study and may be used in conjunction with the improved Kelbg form for electron-electron interactions from Ref.~\onlinecite{Filinov04} or with the electron-electron interactions discussed in Section~\ref{QSP}.

\section{Molecular Dynamics simulations of Carbon Plasmas}
\label{MDCarbon}

Our molecular dynamics simulations were carried out in the canonical ensemble using 14,000 particles (2,000 carbon ions and 12,000 electrons, corresponding to full ionization) in a periodically replicated cell with dimensions chosen according to the desired mass density using the massively-parallel ddcMD code.  The time step in the simulations was 
between $1\times 10^{-5}$ and $5\times 10^{-5}$~fs, and was chosen based on convergence studies of the total energy.  We applied a Nos\'{e}-Hoover thermostat to maintain constant temperature.  Because these simulations were focused solely on a 
calculation of pressure, internal energy, and static pair correlation functions, we applied mass scaling to improve the efficiency of the simulations, setting $m_C=m_e$ in the kinetic energy term of Newton's equation of motion.  The statistical potentials and concomitant force terms in Newton's equation were still computed using the actual masses of both the carbon nucleus and the electron.

Table~\ref{carbonresults} provides the computed internal energies and pressures derived from ddcMD simulations of carbon using the fitted improved Kelbg potential and different choices for the 
Pauli interaction.  We also provide the corresponding data from the L9061 equation of state (EOS) model, which is fit to density functional theory, path integral Monte Carlo, 
and available experimental data for carbon.\cite{Benedict14} 
We note that the internal energies tabulated in L9061 are defined relative to the internal energy of a fully bound carbon atom, and we thusly shifted the values to the "true zero" of energy by subtracting 
this contribution from the tabulated values.  
%REV22626
Simulations that failed to converge, marked as "FTC" in Table~\ref{carbonresults}, are discussed in more detail below.

\begin{table*}[t]
\caption{Carbon pressures and internal energies computed via ddcMD for a variety of temperatures and densities using the new improved Kelbg potential and a variety of forms for the Pauli potential.  The fit to the low-density limit of the Lado Pauli potential\cite{Lado} is denoted as "LADO," whereas "DEUTSCH" denotes the Hansen and MacDonald\cite{HansMcD81} form of the potential originally due to Deutsch, Minoo and 
Gombert.\cite{Deutsch78a,Deutsch78b}  We do not report values where ddcMD simulations failed to converge (denoted as "FTC") due to the formation of unphysical classical clusters.}
\begin{center}
\begin{tabular}{c|c|c|c|c|c|c}
Pauli Potential &	Temperature &	Density &	ddcMD Pressure &	L9061 Pressure &	ddcMD Internal Energy & L9061 Internal Energy  \\
%QSP model &	Temperature (K) &	Density (g/cc) &	ddcMD Pressure (Mbar) &	L9061 Pressure (Mbar) &	Percent difference in Pressure, ddcMD vs L9061 &	ddcMD Internal Energy (eV/atom)	 & L9061 Internal Energy (eV/atom) &	Percent difference in Internal Energy ddcMD vs L9061	& Plasma Coupling Parameter (\Gamma_{ii})	& Fermi Temperature (K) \\
&(K) & (g/cc) &	(Mbar) &	(Mbar) & (eV/atom)	 & (eV/atom)  \\
\hline
\hline
LADO	&1.16E+05&	0.092&	FTC&	2.19E-01 &	FTC	&7.52E+01\\
LADO&	1.16E+06&	0.092&	4.26E+00&	4.59E+00&	1.74E+03	&1.62E+03\\
LADO&	1.16E+06&	3.500&	1.48E+02&	1.38E+02&	7.29E+02	&9.31E+02\\
LADO&	1.39E+06&	3.500&	1.75E+02&	1.75E+02&	1.13E+03	&1.22E+03\\
NONE&	1.39E+06&	3.500&	1.65E+02&	1.75E+02&	1.08E+03	&1.22E+03\\
LADO&	1.74E+06&	3.500&	2.28E+02&	2.35E+02&	1.68E+03	&1.65E+03\\
LADO&	2.32E+06&	3.500&	2.76E+02&	3.39E+02&	2.44E+03	&2.34E+03\\
NONE&	2.32E+06&	3.500&	2.67E+02&	3.39E+02&	2.38E+03	&2.34E+03\\
DEUTSCH&	5.80E+06&	3.500&	9.48E+02&	9.92E+02&	5.87E+03&	5.91E+03\\
DEUTSCH&	1.16E+07&	3.500&	1.97E+03&	2.03E+03&	1.12E+04	&1.14E+04\\
NONE&	1.16E+07&	3.500&	1.95E+03&	2.03E+03&	1.12E+04	&1.14E+04\\
LADO&	1.01E+06&	8.493&	FTC&	2.78E+02&	FTC&	6.59E+02\\
LADO&	2.02E+06&	8.493&	6.79E+02&	6.59E+02&	1.79E+03&	1.76E+03\\
NONE&	2.02E+06&	8.493&	6.31E+02&	6.59E+02&	1.69E+03&	1.76E+03\\
LADO&	1.16E+06&	10.000&	FTC&	3.90E+02&	FTC &	7.96E+02\\
DEUTSCH&	1.16E+07&	10.000&	5.54E+03&	5.67E+03&	1.11E+04&	1.11E+04\\
NONE&	1.16E+07&	10.000&	5.50E+03&	5.67E+03&	1.10E+04&	1.11E+04\\
\end{tabular}
\end{center}
\label{carbonresults}
\end{table*}%

%REV22626
Table~\ref{cresults2} lists the percent differences in pressure and internal energy between the ddcMD results and the L9061 table, computed relative to the tabular values in L9061.  Generally, we find that the QSPs provide internal energy results that are within better than +/-10\% of L9061 for conditions where the electrons are non-degenerate and the K-shell is at least partially ionized.  At temperatures where the K-shell is expected to be $\gtrsim 50$\% occupied, we find that the QSPs produce unphysical classical recombination and clustering in the ddcMD simulations.  The computed pressures show similarly reasonable agreement with the 
L9061 table in situations where the electrons are non-degenerate and the K-shell is at least partially ionized, with the exception of one anomalous point at 
$T=2.32\times 10^6$~K and $\rho=3.5$~g/cc.  This particular point bears further investigation, though our present hypothesis is that this is an anomaly in the 
ionic component of the L9061 equation of state.  We also find that the inclusion of the Pauli potential generally improves 
agreement with the L9061 model relative to simulations without a Pauli potential for most temperatures.  For the highest temperatures studied, which are approaching the ideal gas limit where the thermal energy is much greater than the interaction energy of the system,
the Pauli potential has little impact on the computed EOS as expected.

\begin{table*}[t]
\caption{Summary of differences between computed pressures and internal energies from the ddcMD simulations and L9061.  The percent differences are computed relative to the tabular values in L9061, as listed in Table~\ref{carbonresults}.  Values of the degeneracy parameter, $\tau=T/T_F$, are also listed for each simulation.}
\begin{center}
\begin{tabular}{c|c|c|c|c|c}
Pauli Potential &	Temperature &	Density &	Pressure  &	Internal Energy  & $\tau$ \\
& & & Difference & Difference & \\
%QSP model &	Temperature (K) &	Density (g/cc) &	ddcMD Pressure (Mbar) &	L9061 Pressure (Mbar) &	Percent difference in Pressure, ddcMD vs L9061 &	ddcMD Internal Energy (eV/atom)	 & L9061 Internal Energy (eV/atom) &	Percent difference in Internal Energy ddcMD vs L9061	& Plasma Coupling Parameter (\Gamma_{ii})	& Fermi Temperature (K) \\
&(K) & (g/cc) &	(Percent)  & (Percent) & (Dimensionless) \\
\hline
\hline
LADO&	1.16E+06&	0.092&	-7.19 & 7.37 & 29.9\\
LADO&	1.16E+06&	3.500&	7.33& -21.7 & 2.65\\
LADO&	1.39E+06&	3.500&	0.08&	-7.26 & 3.18\\
NONE&	1.39E+06&	3.500&	-5.71&	-11.45 & 3.18\\
LADO&	1.74E+06&	3.500&	-2.85&	1.38 & 3.98\\
LADO&	2.32E+06&	3.500&	-18.7&	4.16 & 5.30\\
NONE&	2.32E+06&	3.500&	-21.3&	1.72 & 5.30\\
DEUTSCH&	5.80E+06&	3.500&	-4.44&	-0.631 & 13.2\\
DEUTSCH&	1.16E+07&	3.500&	-3.16&	-1.43 & 26.5\\
NONE&	1.16E+07&	3.500&	-4.00&	-1.36 & 26.5\\
LADO&	2.02E+06&	8.493&	3.05&	1.76 & 2.55\\
NONE&	2.02E+06&	8.493&	-4.22&	-4.27 & 2.55\\
DEUTSCH&	1.16E+07&	10.000&	-2.24& -0.219 & 13.2\\
NONE&	1.16E+07&	10.000&	-3.05&	-0.379 & 13.2\\
\end{tabular}
\end{center}
\label{cresults2}
\end{table*}%

\begin{figure}
\centerline{
\rotatebox{-90}{\resizebox{2.5in}{!}{\includegraphics{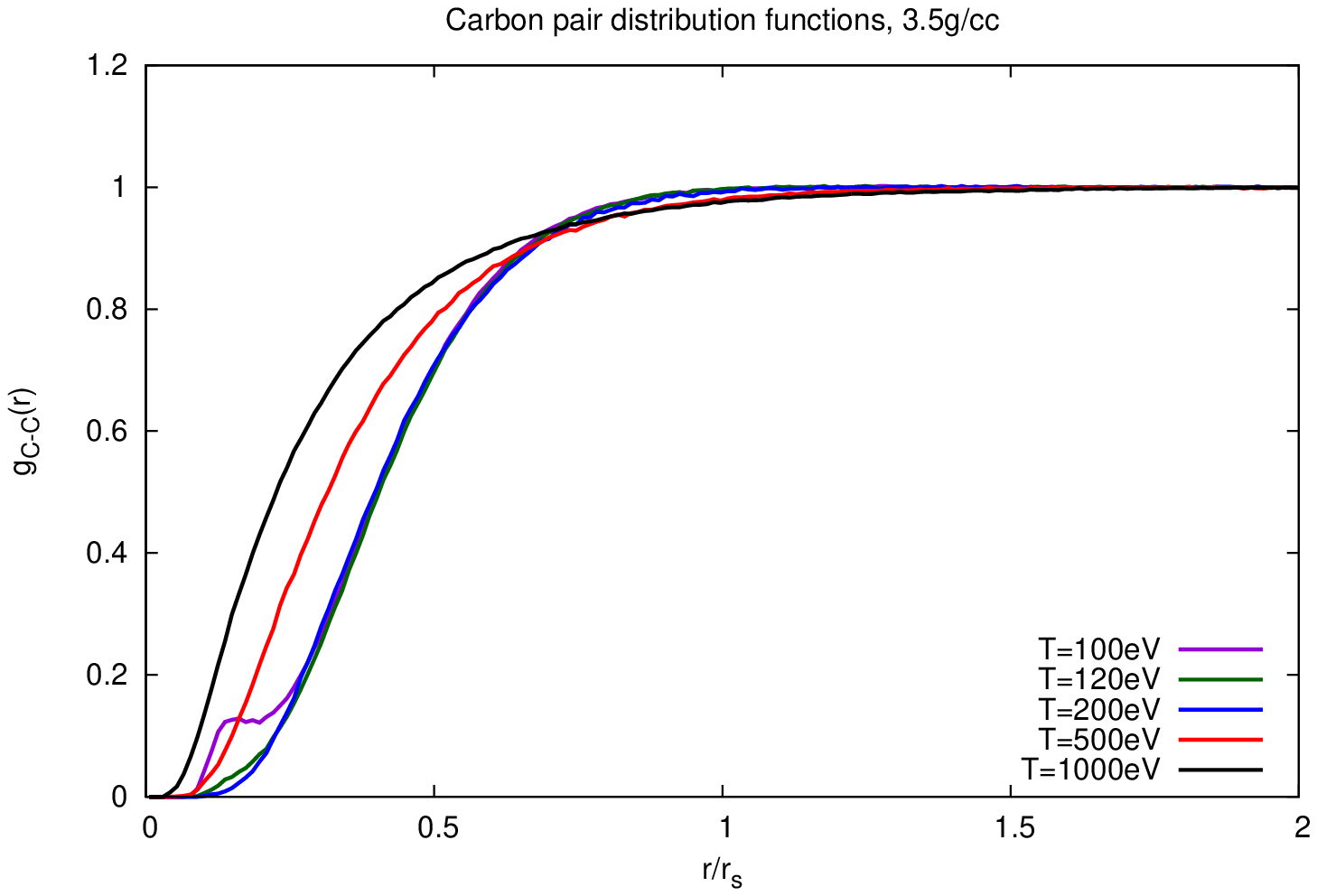}}}}
\caption{Simulated ion-ion pair distribution functions for carbon at 3.5g/cc from ddcMD simulations at a variety of temperatures plotted as a function of distance relative to the ion sphere radius defined as $r_s=\left(\frac{3}{4\pi\rho_i}\right)^{1/3}$ where $\rho_i$ is the ion number density.  We note that 1~eV is equivalent to $1.1604 \times 10^4$~K.}
\label{CCgofr}
\end{figure}

In order to understand why certain simulations did not converge, we also examined the radial pair distribution functions for the ions. Figure~\ref{CCgofr} shows the ion-ion pair distribution function computed in the ddcMD 
simulations for several temperatures along the 3.5~g/cc isochore plotted as a function of the inter-particle separation divided by the ion sphere radius ($r/r_s=1$ corresponds to the average inter-particle 
separation at this density.)  Evidence of unphysical classical carbon clusters is demonstrated in the pair distribution by the appearance of an apparent peak around $r/r_s < 0.5$ at $T=100$~eV and $T=120$~eV ($1.39 \times 10^6$~K and $1.74\times 10^6$~K).   The lack of the 
off-diagonal terms of the quantum density matrix limits the ability of this classical description to incorporate Pauli exclusion, diffraction, and interference that are necessary to provide a more realistic description of the electronic structure as recombination becomes energetically favorable in the carbon plasma.  Interestingly, we 
generally found better stability in the ddcMD simulations with these potentials at low temperatures than in hypernetted-chain simulations, likely due to better numerical stability in the MD approach, possibly introduced by a more adequate treatment of three-body and higher terms in the MD.  We found that hypernetted-chain simulations along the 3.5~g/cc isochore failed 
to converge for $T< 500$~eV ($5.80\times 10^6$~K) due to the presence of attractive electron-ion interactions, which led to stiff solutions that prevented the convergence of our iterative algorithm based on the work of Ng.\cite{Ng74}  Eventually, when the temperature is low enough or the density is high enough, the molecular dynamics simulations also fail to converge within reasonable simulation times, similarly showing the formation of unphysical clusters. 

\begin{figure}
\centerline{
\rotatebox{-90}{\resizebox{2.5in}{!}{\includegraphics{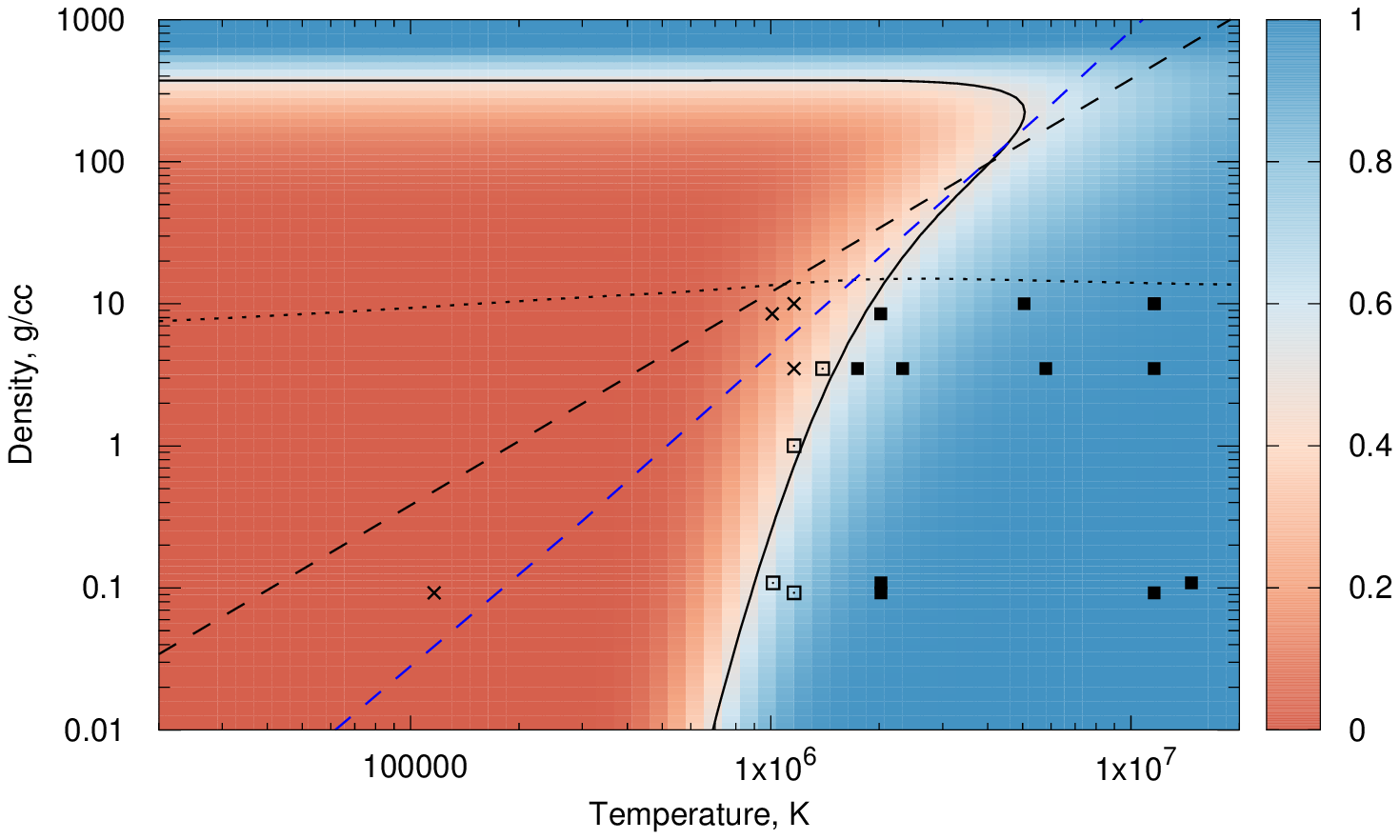}}}}
\caption{Qualitative assessment of the validity of the fitted improved Kelbg potential for carbon as a function of density in grams per cubic centimeter and temperature in Kelvin.  Colormap shows the relative K-shell occupation on a scale of 0 (completely occupied) to 1 (completely unoccupied) based on the computed $1s$ orbital energy and electron chemical potential from Purgatorio.  The solid black contour line corresponds to 50\% occupation of the K-shell, the dotted black line corresponds to the density-temperature track for the principal Hugoniot of diamond from L9061, the dashed black line corresponds to $T=T_F$, and the dashed blue line corresponds to the boundary set by the equality in Eq.~\ref{validline} assuming $Z=6$.  Filled squares denote densities and temperatures for which the ddcMD simulations with QSPs were fully converged and did not exhibit unphysical clustering. The crosses indicate densities and temperatures for which the ddcMD calculations did not converge and the open squares indicate densities and temperatures where the internal energy and pressure were converged but evidence of unphysical quasi-bound states between carbon ions was observed.}
\label{colormapPurg}
\end{figure}

Figure~\ref{colormapPurg} shows a summary of these results for carbon.  The underlying color map shows the relative population of the K-shell, ranging from completely unpopulated (blue) to fully filled (red), 
according to  Purgatorio, which is a fully quantum 
average-atom density functional theory method that includes relativistic effects and incorporates both the change in energy of the K-shell and the pressure ionization of the plasma through the atom-in-jellium approximation.\cite{Purg}
As previously noted by other authors,\cite{murillo2013,TD25} the definition of an effective ionic 
charge $\bar{Z}$ is somewhat arbitrary and dependent upon choices regarding ion sphere radius.  We computed the population of the K-shell based on the chemical potential from Purgatorio and the actual 
energy of the $1s$ state in the average atom description, consistent with other recent studies.\cite{Zhang2018,Zhang2019,Zhang2020,militzerAl}   The solid black contour line shows the conditions at  which we thusly compute 50\% occupation of the K-
shell, while the dotted line shows the principal Hugoniot for diamond, according to L9061, and the dashed black line shows the conditions where the plasma temperature is equal to the Fermi temperature, $T_F$, with conditions to the right of the line corresponding to $T>T_F$.  The blue dashed line indicates the conditions which meet the equality in Eq.~\ref{validline}, with conditions to the right of the line corresponding to those where it is anticipated that an explicitly three (or more)-body solution of the quantum problem is not required in the determination of the diffractive QSP assuming full ionization of the carbon ($Z=6$).    The ddcMD 
calculations are shown on the plot as filled squares for calculations that have yielded internal energies within 10\% agreement with L9061 and where no unphysical classical clusters are observed, open squares for cases where results are 
close to L9061 but evidence of unphysical clusters is observed in the 
radial pair distributions, and as an "x" for conditions where the calculation fails to converge due to unphysical clustering.  

Comparing Fig.~\ref{colormapPurg} to Fig.~6 from Ref.~\onlinecite{Barker71}, we find that the 
agreement/disagreement between ddcMD and L9061 essentially follows the region of validity predicted by Barker when the Purgatorio ionization of the K-shell is used as the baseline for comparison.  
We also note that the underlying electron thermal model for the L9061 equation of state table is based on Purgatorio as a point of consistency for this comparison between the theoretical models.  For most of the cases where the ddcMD internal energies agree with the L9061 table we also find that the pressures computed via ddcMD agree with L9061.  In the cases studied here, we found one exception on the 3.5 g/cc isochore at $T= 2.32 \times 10^6$~K ($T=200$~eV).  

The key finding of these comparisons is that the QSPs can apparently provide a reasonable description of the EOS of the carbon plasma across a broad range of conditions, but that the efficacy of these potentials breaks down as the K-shell fills. This approach could therefore be used as a high efficiency method for examining EOS at 
%REV22626
high temperature, particularly as a point of comparison with 
restricted PIMC simulations. In fact, the differences in pressure reported in this study are similar in magnitude with those reported in comparisons between 
restricted PIMC and several density functional theory approaches in recent EOS development studies.\cite{Zhang2018,Zhang2019,Zhang2020}
Furthermore, this method should be applicable to the difficult problem of predicting the high-T EOS of complex mixtures of different elements, provided that each element is sufficiently well ionized.  The breakdown of the QSPs  
 corresponds to the regime at which a more detailed quantum description of the bound state electrons must be incorporated into the electronic structure, and hence where a 
 quantum simulation method is more appropriate.  One way to extend the QSPs to lower temperatures is by making use of an effective charge 
 state ($Z^*$), which only includes the free electrons, to compute the QSPs from the Slater sum, while others have incorporated bound electrons via a modified form of the 
 the Deutsch potential.\cite{Johnson2024}  
 These approaches are similar to the frozen-core approximation in density functional theory and should provide a reasonable description 
 in situations where the subsystem of bound electrons is sufficiently separated, either energetically or spatially, from the 
 effective free electrons so that full quantum exchange and correlation between the bound electrons and free electrons, and/or between the bound electrons associated with multiple nuclei, 
 need not be considered.  In practice, this corresponds to conditions of sufficiently low density and/or high temperature to meet these conditions, which are generally described by Eq.~\ref{validline}.

\section{Conclusion}

In this paper, we have presented a generalized functional form for the fitting parameter in the improved Kelbg potential that is applicable for $Z>1$.  We also examined the viability of 
computing the equation of state (EOS) of carbon using the improved Kelbg potential in a molecular dynamics simulation as a test of the validity of the potentials.  
Generally, we find good agreement of the molecular dynamics calculations with an equation of state model 
that is fit to quantum simulations (L9061) at conditions where the K-shell of carbon is close to fully ionized.  
Our study suggests that this method could be leveraged to provide calculated EOS values 
more efficiently than the path integral Monte Carlo approach in the very high temperature regime, in addition to providing 
extended a unique capability for: 1) studying finite-size effects in  
quantum-based simulation methods, 2) addressing mixtures with low concentrations of one or more species, and 3) 
examining the impacts of many-body dynamics and correlations in dilute plasma regimes.\cite{devriendt2024classical}  
However, the applicability of this method is generally limited to cases where the thermal de Br\"{o}glie wavelength is small compared to typical inter-particle length scales in the problem so that a full 
many-body solution to the quantum problem is not required to capture the thermodynamic properties of the system.  
The analytic form of the improved Kelbg potential presented here provides a pathway for practical and efficient implementation in standard codes, such as the Sarkas molecular dynamics code\cite{sarkas} or LAMMPS,\cite{LAMMPS} providing a resource for others in the community. Future improvements to these potentials could be made by developing means of extracting effective pair potentials from  
path integral Monte Carlo simulations, such as what was done in Ref.~\onlinecite{TD22}, however care must be taken to ensure that the impacts of many-body interactions are not double-counted when developing potentials for use in explicit many body classical MD or MC simulations.

\begin{acknowledgments}
This work was performed under the auspices of the U.S. Department of Energy by Lawrence Livermore National Laboratory under Contract No. DE-AC52-07NA27344.  Parts of this work were 
funded by the Laboratory Directed Research and Development Program at LLNL under project tracking code 09-SI-011 and 12-SI-005.  
H.~D. Whitley is grateful to the DOE for support provided through a PECASE Award and ASC-PEM.

\end{acknowledgments}

%\appendix

%\section{Appendixes}

\section{References}

\bibliographystyle{aipnum4-2}
%\bibliography{sp2_dec12}% Produces the bibliography via BibTeX.
%aipnum4-2.bst 2019-01-14 (MD) hand-edited version of apsrev4-1.bst

\end{document}